\documentclass[reprint,superscriptaddress,twocolumn,aps,prb,showpacs]{revtex4-1}
\usepackage{graphicx,float}
\usepackage{amsbsy,amssymb,amsfonts,amsmath}
\usepackage{url}

\newcommand{\g}{pdf}

\begin{document}
\title{Coulomb Glasses: A Comparison Between Mean
Field and Monte Carlo Results}
\author{E. Bardalen}
\affiliation{Department of Physics, University of Oslo, P.O.Box 1048
  Blindern, N-0316 Oslo, Norway}
\author{J. Bergli}
\email[]{jbergli@fys.uio.no}
\affiliation{Department of Physics, University of Oslo, P.O.Box 1048
  Blindern, N-0316 Oslo, Norway}
\author{Y. M. Galperin}
\affiliation{Department of Physics, University of Oslo, P.O.Box 1048
  Blindern, N-0316 Oslo, Norway}
\affiliation{A. F. Ioffe Physico-Technical Institute RAS, 194021 St. Petersburg, Russian Federation}
\affiliation{Centre for Advanced Study at the Norwegian Academy of Science and Letters,
0271 Oslo, Norway.}

\begin{abstract}
Recently a local mean field theory for both eqiulibrium and transport
properties of the Coulomb glass was proposed [A. Amir  \textit{et al.}, Phys. Rev. B {\bf 77},
165207 (2008); {\bf 80}, 245214 (2009)]. We compare the
predictions of this theory to the results of dynamic Monte Carlo
simulations. In a thermal equilibrium state we compare the density of
states and the occupation probabilities. We also study the transition
rates between different states and find that the mean field rates
used in the aforementioned papers 
underestimate a certain class of important transitions. We propose
modified rates to be used in the mean field approach which take into
account correlations at the minimal level in the sense that
transitions are only to take place from an occupied to an empty
site. We show that this modification accounts for most of the
difference between the mean field and Monte Carlo rates. The linear
response conductance is shown to exhibit the Efros-Shklovskii
behavior in both the mean field and Monte Carlo approaches, but the
local mean field method strongly underestimates the current at low
temperatures. When using the modified rates better agreement is
achieved.
\end{abstract}

\maketitle

\section{Introduction}

Electronic states in disordered materials can be localized and at zero
temperature the material can be insulating.  At finite temperature,
transport can still occur trough hopping between the localized
stated. The energy mismatch between the states must be supplied by
emission or absorption of phonons, so the process is called
phonon-assisted hopping. This has been studied for many years, one of
the early major developments being Mott's concept of variable range
hopping\cite{mott} (VRH) leading to the Mott law for the temperature
dependence of the conductivity:
\begin{equation}
\sigma \propto \exp[-(\tilde{T}_0/T)^{1/(d+1)}]\, .
\end{equation}
Here $T$ is temperature, $\tilde{T}_0$ is some constant dependent on
the material parameters while $d$ is the dimensionality of the
conduction problem. Mott gave only a heuristic derivation of this law
based on the idea of a competition between tunneling distance and
energy difference. This was later rigorously derived using percolation
arguments.\cite{ambegaokar,ES71,Pollak72,Shklovskii72} This puts the theory on a firm basis but
it applies only in the case of non-interacting electrons. If Coulomb
interactions are important, the theory has to be extended. The first
step in this direction was the understanding of the 
interaction-induced dip
 in the
density of states (DOS).\cite{Pollak70,Pollak71,Srinivasan} 
Based on a stability argument on states which are stable to all
one-particle jumps, Efros and Shklovskii\cite{ES75}  derived
 an upper bound on the density of states increasing as $\epsilon^{d-1}$ where the excitation energy
$\epsilon$ is counted from the Fermi level. 
At finite temperatures the
gap is smeared by thermal fluctuations.\cite{PikusEfros} 

While the
arguments up to this point seem rigorous, one then proceeds with some
type of mean field (MF) treatment.\cite{ES} Since the energy of a
certain site depends on the occupancy of the other sites through the
Coulomb interaction, the site energies will fluctuate in time as jumps
take place. Instead of following the fluctuatig occupation numbers,
one replaces them with some average occupancy of each site. In this
way, one can repeat Mott's argument replacing the uniform density of
states with the Coulomb gap form, leading to the Efros-Shklovskii (ES)
law for conductivity:\cite{ES75}
\begin{equation}\label{eq:ES}
\sigma \propto \exp[-(T_0/T)^{1/2}]\, .
\end{equation}
While this law has been seen in many experiments, it is difficult to prove rigorously that
a manifestly many-particle concept like the Coulomb gap density of states 
can be used in a single particle picture like that given by Mott.

Recently there was renewed interest in these questions, and local mean
field (LMF) theory  was revisited\cite{amirMF} 
and applied to time-dependent phenomena, see
Refs.~\onlinecite{AmirAnu,*AmirPNAS} for reviews. 
Having been applied  to the VRH problem\cite{amirVRH} the LMF theory has reproduced the ES 
temperature dependence  (\ref{eq:ES}) of the conductance. 
The LMF approach differs form the conventional one,~\cite{ES} which we refer to as
ESMF, by two aspects. Firstly, the LMF equations solved numerically at finite temperature
are intended to represent equilibrium state and  therefore to allow for  the finite-temperature 
smearing of the Coulomb gap. Contrary,  the ESMF approach uses the ground state as a starting point and allows for 
finite temperatures only in the occupation numbers. Secondly, the LMF uses a different
expression for the transition rate between the states, which neglects Coulomb correlations in the transferred energy
that seems to be inconsistent. 
This issue will be discussed in detail  in Sec.~\ref{sec:MF}. Because of that, the LMF approach 
is still beyond full control, and there are still questions about its validity.  It is
therefore important to check the results, see if they can be trusted
in some regions of parameters and in this way obtain limits of the
validity of the mean field approximation. In this work we compare the
MF analysis to dynamic Monte Carlo simulations of the hopping
process. This allows us to compare both equilibrium properties like
the DOS and the occupation numbers, as well as dynamical aspects like
the transition rates and the conductance.

We will conclude that the LMF  strongly underestimates the rate of a
certain important class of transitions, and this leads to an
underestimation of the current. We propose a modification of the
transition rates entering the mean-filed scheme
in the spirit of the considerations of Ref.~\onlinecite{ES}, Secs. 10.1.2 and 10.2.1 
(see also Refs.~\onlinecite{levin82,*levin82a}), which takes into account 
the interaction-induced  correlations at the minimal level, and ensures that a
transition will take place only from an occupied to an empty
site. This gives much better estimates for the transition rates, and
better agreement with the Monte Carlo simulations of the conductance.

Note that recently the question of a finite temperature phase transition to a low temperature glass phase was discussed, and it was found that mean field theories 
\cite{pastor99,*vojta93,*muller04,*muller07,*pankov05} predict such a phase transition while this was not found in Monte Carlo simulations.\cite{katzgraber09, goethe2009}

The paper is organized as follows. In Sec. \ref{sec:MF} we recall the
MF approach and explain our modified transition rates. In
Sec. \ref{sec:numerics} we outline the numerical Monte Carlo method we use
to test the MF results.  We  comparw the properties (DOS and
occupation probabilities) of equilibrium states in
Sec. \ref{sec:equilibrium}. The transition rates are compared in
Sec. \ref{sec:rates} and the conductance in Sec. \ref{sec:cond}. In
Sec. \ref{sec:discussion} we summarize the results.

\section{Mean field equations}\label{sec:MF}

In Ref.~\onlinecite{amirMF}, a 
LMF approach was developed
and then applied to the calculation of the conductivity in the
variable range hopping regime.\cite{amirVRH} We give here a brief
summary of their method. We model the system as a set of sites, each
of which can be either empty or occupied by one electron. For
numerical convenience, the sites are arranged in a two dimensional
square lattice
with lattice constand $d$.
Each site is given an energy $\phi_i$ 
uniformly distributed
in the interval $[-U,U]$. The total single-particle (addition or
subtraction) energy (SPE) of site $i$ is then
\begin{equation}
\epsilon_i = \phi_i + \sum_{j\neq i}\frac{n_j-\nu}{r_{ij}} \, .
\end{equation}
Here distances $r_{ij}$ between sites $i$ and $j$ are measured in
units of $d$ while energies are measured in units of $e^2/d$ where $e$
is the electronic charge.  Background dielectric constant $\kappa$ is set to 1. 
$n_j=0,1$ is the occupancy of site $j$.
The compensating background charge $\nu$ measured in unites of $e$ and
associated to each site is introduced to keep the system neutral. We
have considered the case of half filling, so that the number of
electrons is half the number of sites, and therefore $\nu= 1/2$. The
occupation numbers $n_i$, and therefore also the energies
$\epsilon_i$, fluctuate in time as the electrons jump between the
sites. In the MF approximation, one replaces the fluctuating
quantities by their averages, associating to each site an
average occupancy $f_i = \langle n_i\rangle$ and average energy
\begin{equation}\label{MFE}
E_i = \langle\epsilon_i\rangle = \phi_i + \sum_{j\neq i}\frac{f_j-\nu}{r_{ij}}\, .
\end{equation}
The average occupancy is postulated to be given by the Fermi distribution
at the average energy, 
\begin{equation}\label{MFF}
f_i = f_{\text{FD}}(E_i) \equiv  \left(e^{E_i/T} + 1\right)^{-1}.
\end{equation}
Equations \eqref{MFE} and \eqref{MFF} form a closed set, which we call
the mean field equations.  It has been shown\cite{amirVRH} that the
solutions of these equations give a density of states with a linear
(in 2 dimensions) gap at the Fermi level as expected from the analysis
of the Coulomb gap by the stability condition in the ground state.\cite{ES}

To calculate the conductance we must
consider the transition rates between the different sites. 
If an electron hops from site $i$ to site $j$ the change in 
the systems
energy is 
\begin{equation}\label{spe}
\Delta \epsilon_{ij} = \epsilon_j-\epsilon_i- 1/r_{ij}\, . 
\end{equation}
Here the final term arises because in the definition of $\epsilon_j$
it was assumed that site $i$ was initially occupied and site $j$ was
initially empty. Therefore the hopping event creates an electron on
site $j$ and a hole on site $i$, which attract each other with the
energy $r_{ij}^{-1}$.  The energy change $\Delta \epsilon_{ij}$ must
be supplied by the emission or absorption of a phonon. The rate of
such a process is given by\cite{ES}
\begin{equation}\label{Gamma}                    
\Gamma_{ij} = \tau_0^{-1} (|\Delta \epsilon_{ij}|/E_0) e^{-2 r_{ij}/a}|N(\Delta \epsilon_{ij})|n_i(1-n_j)
\end{equation} 
where $E_0$ is some energy scale characteristic of the electron-phonon
interaction (we set it to 1 in the following),  $\tau_0$ is a
microscopic timescale which we take as our unit of time,  $a$ is the localization length,
which we set equal to the lattice constant $d$,
and  $N(E) = 1/(e^{E/T}-1)$ is the 
Planck function. The difference $\Delta \epsilon_{ij} > 0$ corresponds
 to phonon absorption, while $\Delta \epsilon_{ij} < 0$ corresponds to phonon emission. 
In this case, $|N(\Delta \epsilon_{ij})|=
N(|\Delta \epsilon_{ij}|)+1$ allowing for spontaneous emission.

In the mean field approximation the product $|N(\Delta
\epsilon_{ij})|n_i(1-n_j)$ is replaced by its ensemble average,
$\langle |N(\Delta \epsilon_{ij})|n_i(1-n_j) \rangle$, which is then
decopled into a product of averages, $|N(\langle \Delta
\epsilon_{ij}\rangle)| f_i(1-f_j)$. As a result, the rates
$\Gamma_{ij}$ are replaced by
\begin{equation} \label{gamma}
\gamma_{ij} = \frac{1}{\tau_0} \frac{|\langle\Delta
  \epsilon_{ij}\rangle|}{E_0}e^{-2 r_{ij}/a}|N(\langle\Delta
  \epsilon_{ij}\rangle)|f_i(1-f_j).
\end{equation}
In Ref.~\onlinecite{amirVRH} it was argued
that the energy change should be taken as the difference in the
average energies 
\textit{without including the last term} in Eq.~\eqref{spe},
\begin{equation}\label{deApprox}
 \langle\Delta \epsilon_{ij}\rangle = E_j -E_i \, .
\end{equation}
The formal reason for omitting the self-interaction term is that it is
necessary in order to have detailed balance in equilibrium,
$\gamma_{ij}^0 = \gamma_{ji}^0$ where the superscript 0 indicates that
these are the rates in equilibrium in the absence of  an applied
electric field. It is also argued that this is natural in a mean field
approach since charge can be thought of as transferred continuously and
this term is proportional to the transferred charge squared which
is then infinitesimal. 

The above considerations seem worrying for two (closely related)
reasons. First, the energies of the phonons must be calculated
including this term, so it appears that the mean field approach will
give an incorrect energy balance between the electrons and the phonon
bath. Second, there will be a number of transitions which are given a
mean field rate very different (and much smaller) than the real rate. 
To see this point clearly, consider the situation where $E_i<0$ and
$E_j>0$ while both $|E_i|\gg T$ and $|E_j|\gg T$. Then $f_i\approx1$
and $f_j\approx0$. This means that it will often be the case that site
$i$ is occupied and site $j$ empty. This is a necessary condition for
the transition from site $i$ to  site $j$ take place, and the fact
that it is likely to happen is reflected in the fact that in
\eqref{gamma} the factor $f_i(1-f_j)\approx1$. But $E_j-E_i$ is
positive and larger than temperature so that the rate is
strongly suppressed by the factor $N(|\langle\Delta
\epsilon_{ij}\rangle|)\approx e^{-(E_j-E_i)/T}$. The real energy
change if the transition is from a  configuration where  site
$i$ is occupied and site $j$ empty and where the single particle
energies $\epsilon_i$ and $\epsilon_j$ are close to the averages $E_i$
and $E_j$ is given by Eq. \eqref{spe}. If the sites are so close that
$\epsilon_j-\epsilon_i<r_{ij}^{-1}$ the process will be an
emission process rather than an absorption process, and $N(|\Delta
\epsilon_{ij}|)\approx1$ so the transition rate is exponentially
larger than what is found using the approximation \eqref{deApprox}. 

It appears that 
for the considered configuration
the approximation \eqref{deApprox} is not a good one
and misses an important physical property of the process. To improve
this while keeping the microscopic balance we propose the following. 
We keep the mean field equations  \eqref{MFE} and \eqref{MFF} for
calculating the average occupation numbers $f_i$. But when calculating
the rate $\gamma_{ij}$ we consider the joint occupation probabilities
for both sites. This follows closely the reasoning of
Ref.~\onlinecite{ES} (Sec 10.1.2) 
except that  there all other
sites were considered to be in the ground state configuration, whereas
here we assume them to have the mean field average occupation. 

Let $F(n_i,n_j)$ denote the probability to find the system with
occupation numbers $n_i$and $n_j$, all other sites having the mean
field occupation $f_k$. $F(n_i,n_j)$ depends on the energy changes
when adding particles to the sites $i$ and $j$. We define
\begin{equation}
E_{i(j)} = \phi_i + \sum_{k\neq i,j}\frac{f_k}{r_{ik}}
\end{equation}
as the energy of site $j$ not taking into account the interaction with
site $j$ (note that here we are omitting the background charge $\nu$
since it will cancel from all energy differences). Starting from the
state where both the sites are empty, the energy changes to the other
states are 
\begin{equation}
 E(n_i,n_j) = E_{i(j)}n_i +E_{j(i)}n_j + n_in_j/r_{ij}.
\end{equation}
We then get 
\begin{equation}
F(n_i,n_j) = Z_{ij}^{-1}e^{-E(n_i,n_j)/T}
\end{equation}
where
$$
 Z_{ij} = 1 + e^{-E_{i(j)}/T}+ e^{-E_{j(i)}/T} + 
 e^{-(E_{i(j)} + E_{j(i)} + 1/r_{ij})/T}
$$ 
is the partition function. The transition rate is then 
\begin{equation}\label{gammaMod}
\gamma_{ij} = \frac{1}{\tau_0} \frac{|\Delta E_{i(j)}|}{E_0}e^{-2
  r_{ij}/a}|N(\Delta E_{i(j)})| F(1,0)\, .
\end{equation}
These rates satisfy the detailed balance condition
$\gamma_{ij}^0 = \gamma_{ji}^0$ in the absence of an applied electric
field and they properly take into account the energy change in the  transition. 
Equation (\ref{gammaMod}) plus a LMF scheme takes into account both  the intra-site and inter-site correlations at the minimal level of ensuring that jumps only take place from occupied to empty sites and the finite temperature modification of the ground state. We refer to this theory as the modified local mean field (MLMF) theory. In the following we will compare this to Monte Carlo simulations.

\section{Numerical algorithm}\label{sec:numerics}

We used the kinetic Monte Carlo algorithm introduced in
Refs.~\onlinecite{tsigankov,tsigankovPRB}. It consists in writing the
rate~\eqref{Gamma} as a product of a distance dependent part,
$\Gamma^D_{ij} \equiv  e^{-2 r_{ij}/a}$, and an energy dependent part, $$\Gamma^E_{ij}
\equiv  \tau_0^{-1} (|\Delta \epsilon_{ij}|/E_0)|N(\Delta
\epsilon_{ij})|n_i(1-n_j)\, .$$
 $\Gamma^D$ is independent of the
configuration and can be precalculated and stored. Since we are
working on a lattice, it depends only on the relative distance, and
this is not too costly. The program first selects an electron at
random and then a possible jump weighted by the rates $\Gamma^D$. If
the final site is empty, the jump is then accepted with probability
$\Gamma^E$. Because of the linear dependence of $\Gamma^E$ on $|\Delta
\epsilon_{ij}|$ it is unbounded for large negative $\Delta
\epsilon_{ij}$. For this reason, the rate has to be limited by some
maximal rate, which we have taken rather arbitrarily as $3T$ where $T$
is the temperature and the rates then normalized so that the maximal
rate has probability 1 for being accepted. We have checked that the
exact value of the maximal rate is not important as long as we are not
applying strong electric fields,\cite{aurora} far beyond the linear
response we will consider below.

\section{Equilibrium states}\label{sec:equilibrium}
\begin{figure}[b]
\begin{center}
\includegraphics[width=0.99\linewidth]{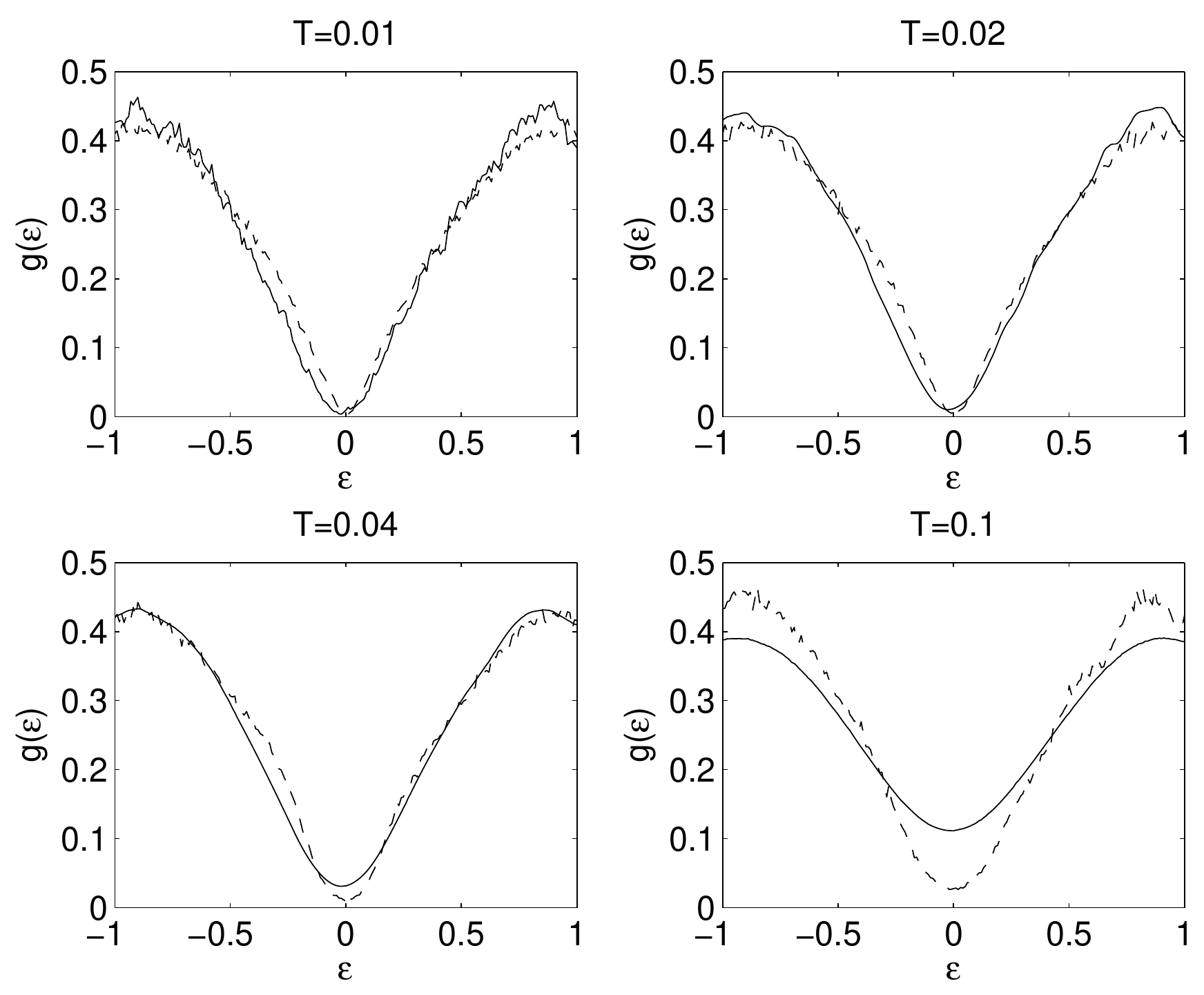}
\end{center}
\caption{\label{fig:CG} The density of states of SPEs at four
temperatures: Solid lines: Averaged Monte Carlo simulations. Dashed lines:
Averaged mean field solutions. For Monte Carlo simulations: Averaging
were done over one ($T = 0.1$), three ($T= 0.04$ ), six ($T= 0.02$)
and six ($T= 0.01$) Monte Carlo simulations;. Mean field solutions
were averaged over 100 solutions for each temperature. }
\end{figure}

We can now compare several properties of the equilibrium states as
described by the LMF and Monte Carlo methods. 
\footnote{Note that MF calculations of DOS do not involve transition rates. 
Therefore, LMF and MLMF calculations of  DOS are equivalent.}
In all comparisons we use the same sample for both methods, so that we are
certain that all differences are a result of the approximations of the
mean field model. We use a sample of size $100\times100$ sites and an
on-site disorder $U=1$. In
the Monte Carlo simulations we average over from 1 to 6 initial
configurations (at high temperatures, thermal motion is sufficient to
wash out any difference between initial states). From each initial
state we perform $3\cdot10^6$ jumps to equilibrate the system and then
we average over the next $5\cdot10^5$ jumps. We solve the mean field
equations numerically by an iterative procedure. The presented results
are averages over 100 different solutions of the mean field
equations.

Let us first study the density of single particle states ($\epsilon_i$
in the Monte Carlo simulations $E_i$ in the mean field solutions). At
low temperatures this should display the Coulomb gap, which according
to the stability analysis of one-particle stable states should be
linear in two dimensions.\cite{ES} It was previously
shown\cite{glatz2008} that in one-particle stable states with $U=1$ we
get in two dimensions a Coulomb gap which is not fully linear but
rather of the form $\propto|\epsilon|^{1.2}$. With the accuracy that
we are working here we do not expect to see the departure from linear.
For the thermal equilibrium states the corresponding quadratic law in
three dimensions was confirmed.\cite{goethe2009} The mean field
equations was previously shown to give a linear Coulomb
gap\cite{amirMF} at low temperatures. Here we compare the solutions of
the LMF equations to Monte Carlo results at different temperatures
(Fig.~\ref{fig:CG}).

\begin{figure}[b]
\begin{center}
\includegraphics[width=0.99\linewidth]{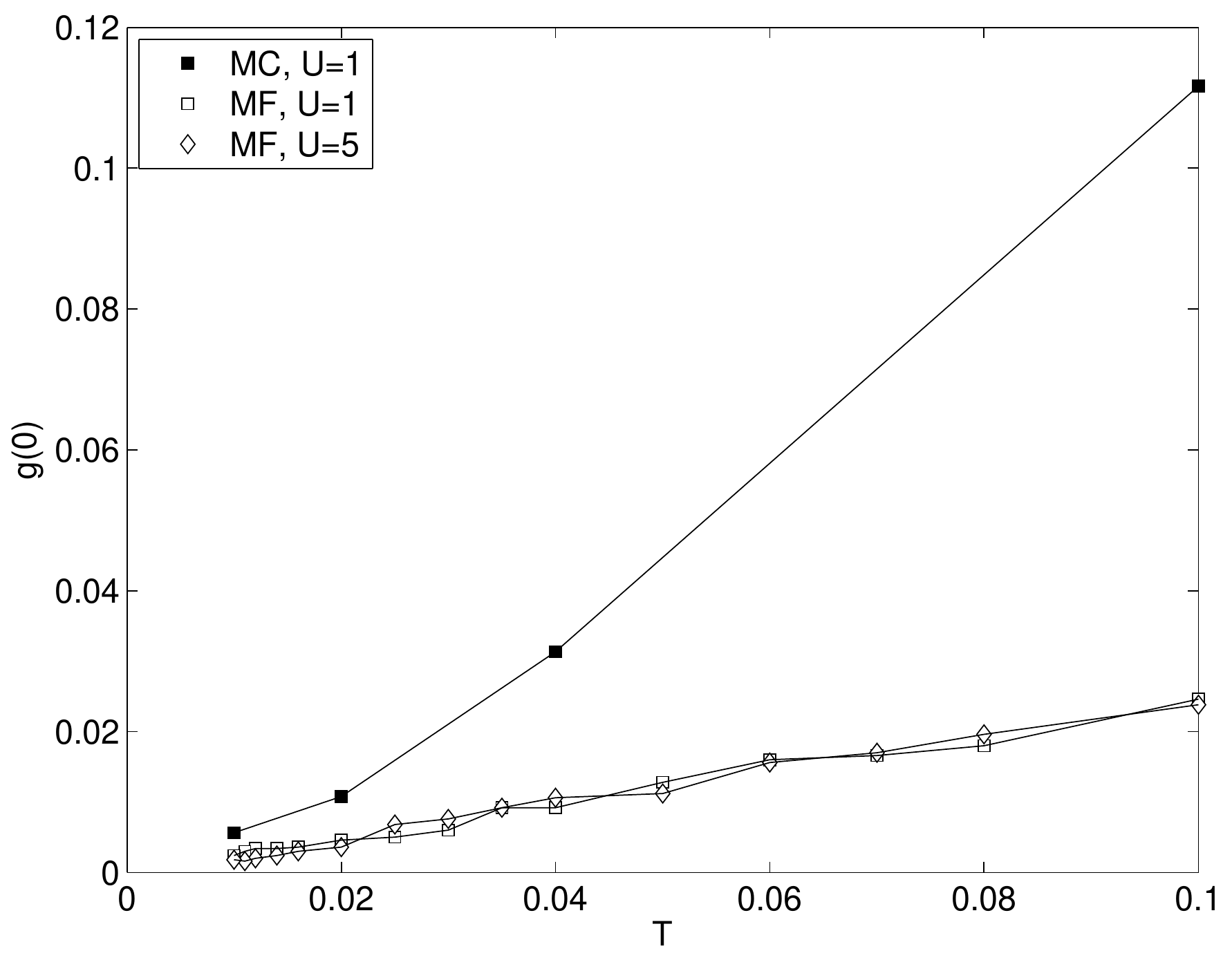}
\end{center}
\caption{\label{fig:g0} Density of states at the Fermi level,
  $g(0)$. The mean field results are the averages over 100 ($U=1$) and
  50 ($U=5$) solutions. Graphs show a near linear dependence on temperature for
  mean field equations, while Monte Carlo solutions are near linear at
  higher temperatures, while showing non-linear behavior at low
  temperatures. }
\end{figure}

At low temperatures we see that the two methods give similar results
and close to the expected linear Coulomb gap.  At higher temperatures
the Coulomb gap is smeared by thermal fluctuations. We observe that
the smearing is much more efficient in the Monte Carlo simulations
than in the mean field solutions.  Thus, the mean field equations
underestimate the smearing of the Coulomb gap at finite
temperatures. This can also be illustrated by plotting the density of
states at the Fermi level, $g(0)$, as function of temperature
(Fig.~\ref{fig:g0}).
In the LMF $g(0)$ is close to a linear function, $g(0)=\alpha T$.  For
 the Monte Carlo we find that the dependence is slightly superlinear,
 but if fitted by a linear function we get a slope of $\alpha=1.54$ in
 reasonable agreement with the previous results,\cite{levin1987} which
 gave $\alpha=1.3$.  For the mean field we get $\alpha=0.25$, which is
 almost twice the result of Ref.~\onlinecite{amirMF} where it was found that $\alpha=0.15$. Note however
 that in that work sites were at random. In addition, the line does not
 pass through the origin as it should, and therefore we do not believe that their result is numerically accurate.
It used $U=5$, but we have  confirmed that we get the same results in that case so this is not
 the source of discrepancy.

A self consistent equation for the energy dependence of the density of states\cite{raikh89} predicts 
the scaling law (in our units) 
$$g(\epsilon, T)= T \mathcal{F}_2 \left( \epsilon/T \right)\, .$$
We find that our LMF  results are compatible with this equation. However, the results of the  Monte Carlo simulations do not collapse to this law. This is also evident from the superlinear temperature dependence shown in Fig. \ref{fig:g0}.

\begin{figure}
\begin{center}
\includegraphics[width=0.99\linewidth]{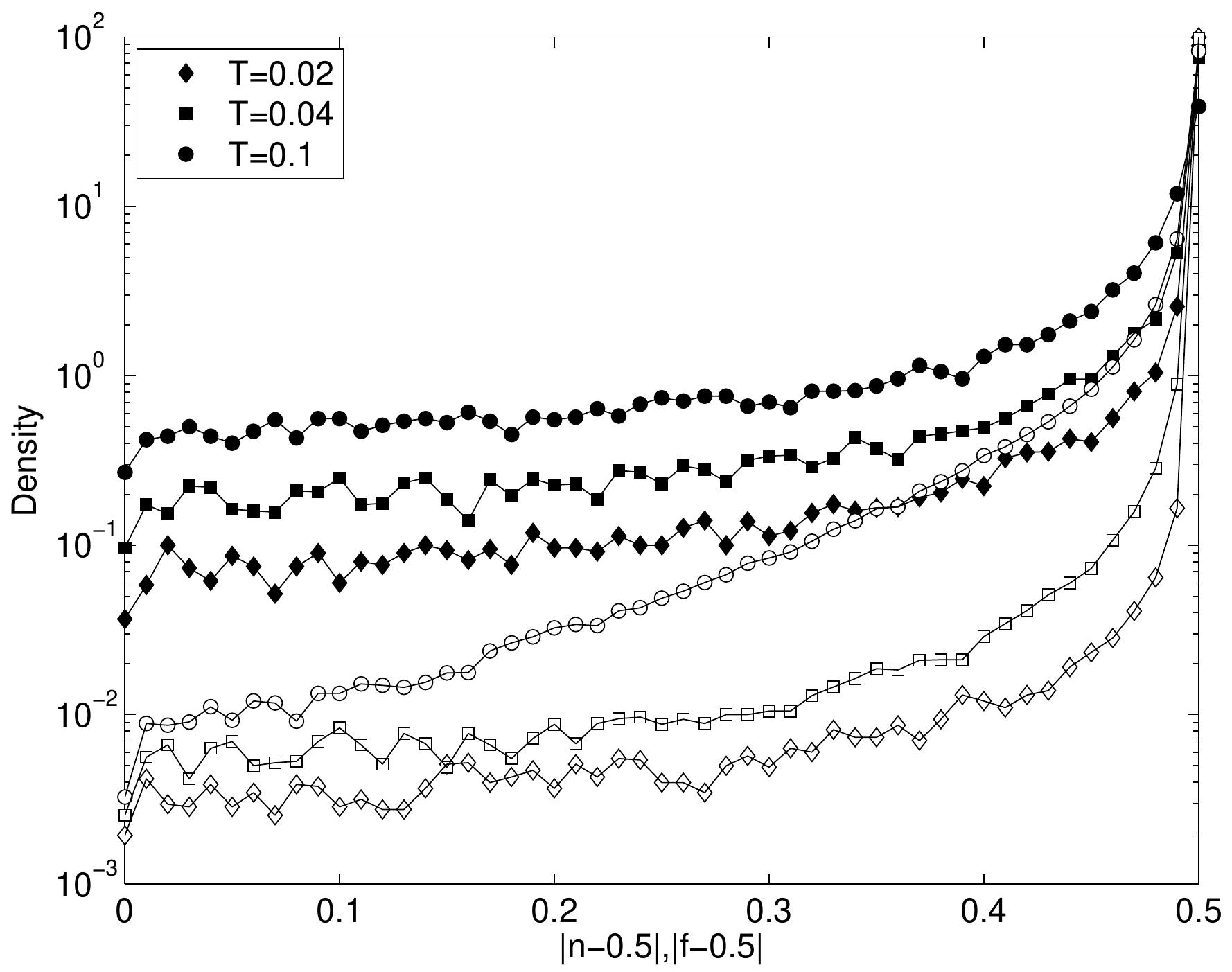}
\end{center}
\caption{\label{fig:n} 
Fraction of sites with intermediate occupation. Empty symbols: Mean field
solution averaged over 100 solutions for each temperature. Filled symbols: Monte Carlo $\langle n_i \rangle$ averaged over $5\cdot 10^5$ jumps.
}
\end{figure}

We can also compare the occupation numbers $f_i$ with the time averaged
occupation numbers $\langle n_i\rangle$ in the Monte Carlo
simulations. Figure~\ref{fig:n} shows the fraction of sites which have
a certain value of $|f_i-0.5|$ or  $|\langle n_i\rangle-0.5|$. 
It is clear that the mean field equations underestimate the number of
sites with intermediate occupation numbers.

\section{Transition rates}\label{sec:rates}

We can compare the transition rates 
calculated using
the different models. In
the Monte Carlo simulations we simply count how often a transition
takes place from a site of energy $E = \epsilon_i$ to a site of energy
$E'=\epsilon_j$. That is, we refer to the energies before the transition
took place. The change in energy is then $\Delta \epsilon =
\epsilon_j-\epsilon_i-1/r_{ij}$.

\begin{figure}[h]
\begin{center}
\includegraphics[width=0.49\linewidth]{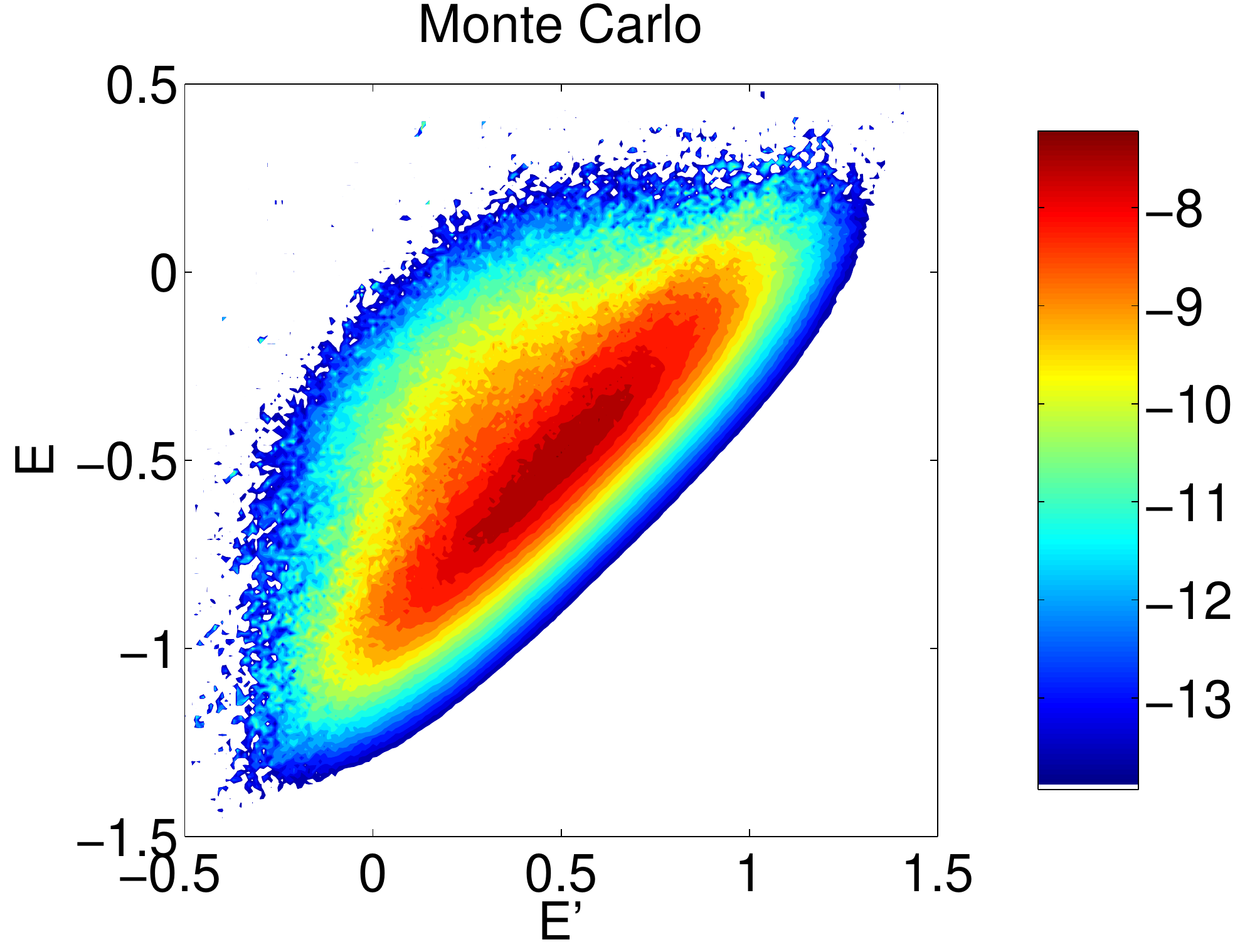} \\
\includegraphics[width=0.49\linewidth]{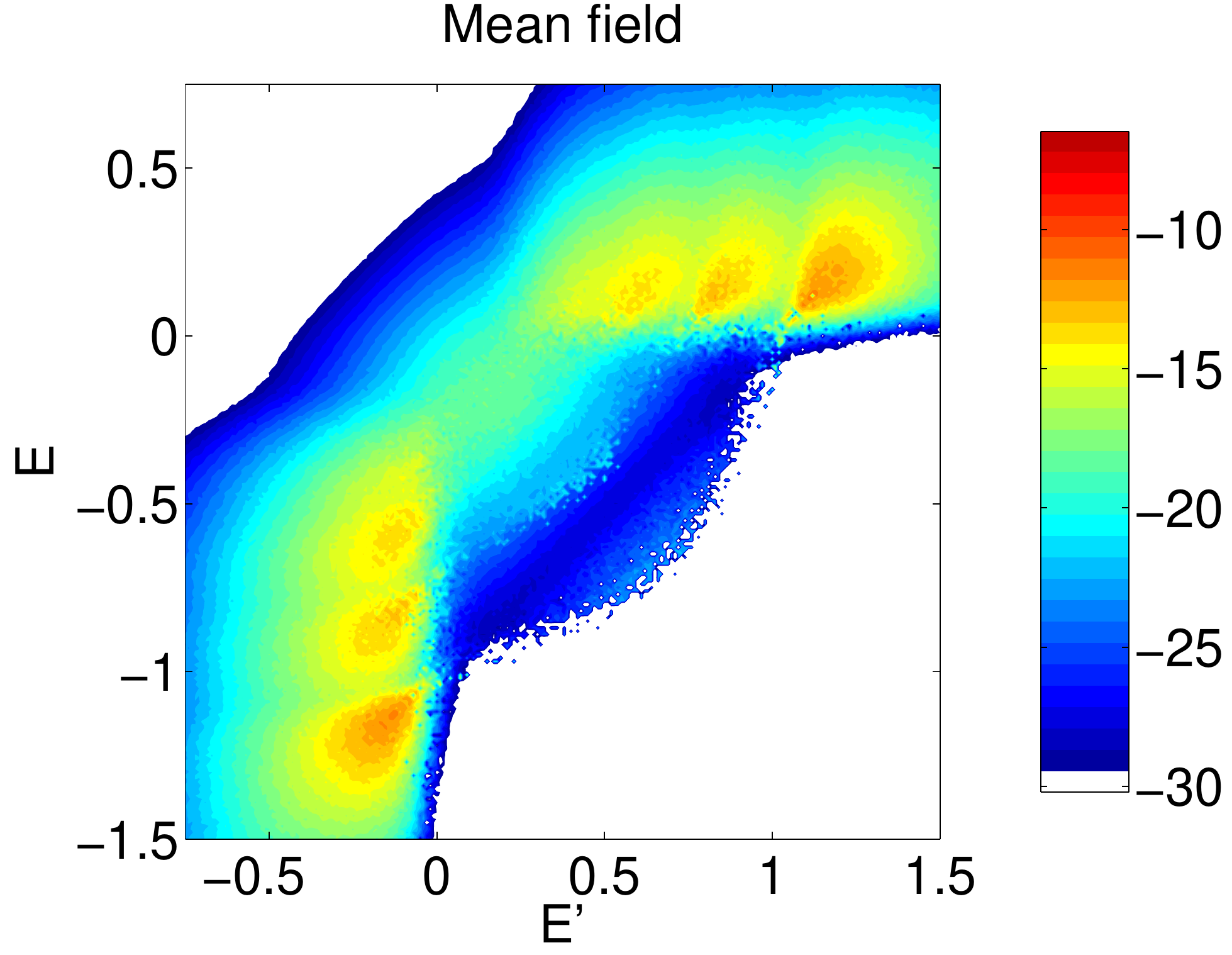}
\includegraphics[width=0.49\linewidth]{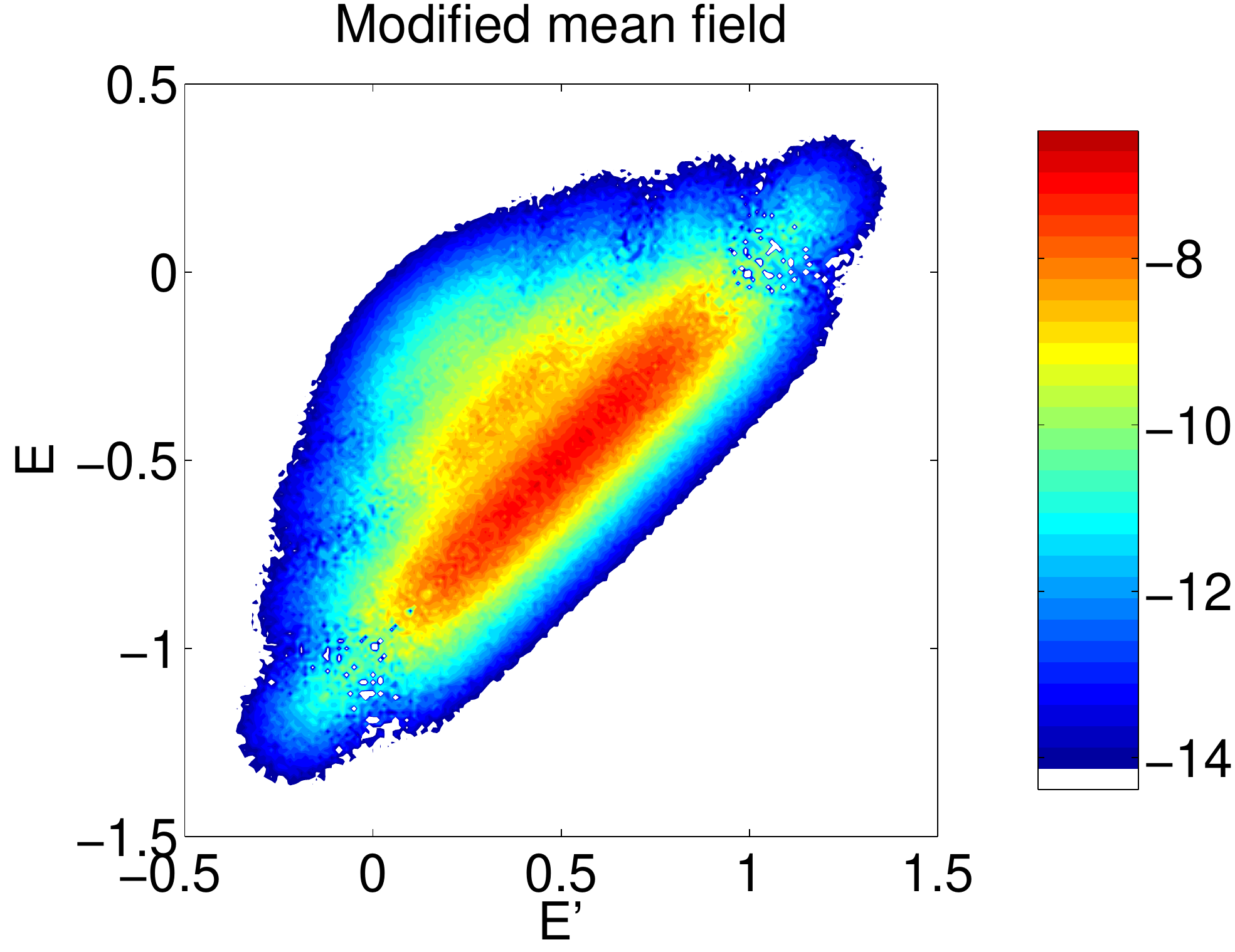}
\end{center}
\caption{\label{fig:rates} Transition rates at T=0.04. Top: Monte
  Carlo. Bottom left: LMF. Bottom right: Modified LMF. For each model
  the logarithm of the average of the sum of all rates from a site of
  energy $E$ to a site of energy $E'$ is shown. White indicates that
  the rate was smaller than the smallest colored rate. }
\end{figure}

In the mean field calculations we have to take some care in
associating the proper energies to a transition so that comparison to
the Monte Carlo results is meaningful. The solution of the mean field
equations provides a set of occupation numbers $\{f_i\}$ and energies
$\{E_i\}$. The energy of site $i$ before the transition is given by
$E=E_{i(j)} = E_i - f_j/r_{ij}$ since this is the energy assuming site
$j$ to be empty. The energy of site $j$ before the transition is
$E'=E_{j(i)}+ 1/r_{ij} = E_j + (1-f_i)/r_{ij}$ since we know that
before the transition site $i$ is occupied. In Fig.~\ref{fig:rates} we
show the average of the sum of the rates from energy $E$ to $E'$ for
the three different models at $T=0.04$. 
We observe clearly what we discussed in Sec. \ref{sec:MF}. The
modified mean field rates closely follow the Monte Carlo results,
while the original mean field rates are smaller for transitions
crossing the Fermi level ($E<0$ and $E'>0$). 

\section{Conductance}\label{sec:cond}

We also calculated the conductance in all three models. In the Monte Carlo
simulations this was done by applying an electric field $E_f$ in the
$x$-direction. This modifies the energy change to $\Delta
\epsilon_{ij}  =\epsilon_j-\epsilon_i-1/r_{ij}- E_f\Delta x_{ij}$. The
transition rates must then be calculated using this energy change but
othervise the simulation is as in the equilibrium case. The current is
measured directly as the transferred charge in the direction of the
field.

In the mean field calculations we find the Miller-Abrahams resistances
\cite{MillerAbrahams}
$ R_{ij} = T/\gamma_{ij}^0$ and construct the resistance network. 
In this case we do not use periodic boundary
conditions in the direction of the field. Rather, we set all sites on
the left edge to one potential and all on the right to a different
potential. We then solve the Kirchhoff equations for the network to
find the current in each resistor. 

\begin{figure}[h]
\begin{center}
\includegraphics[width=0.99\linewidth]{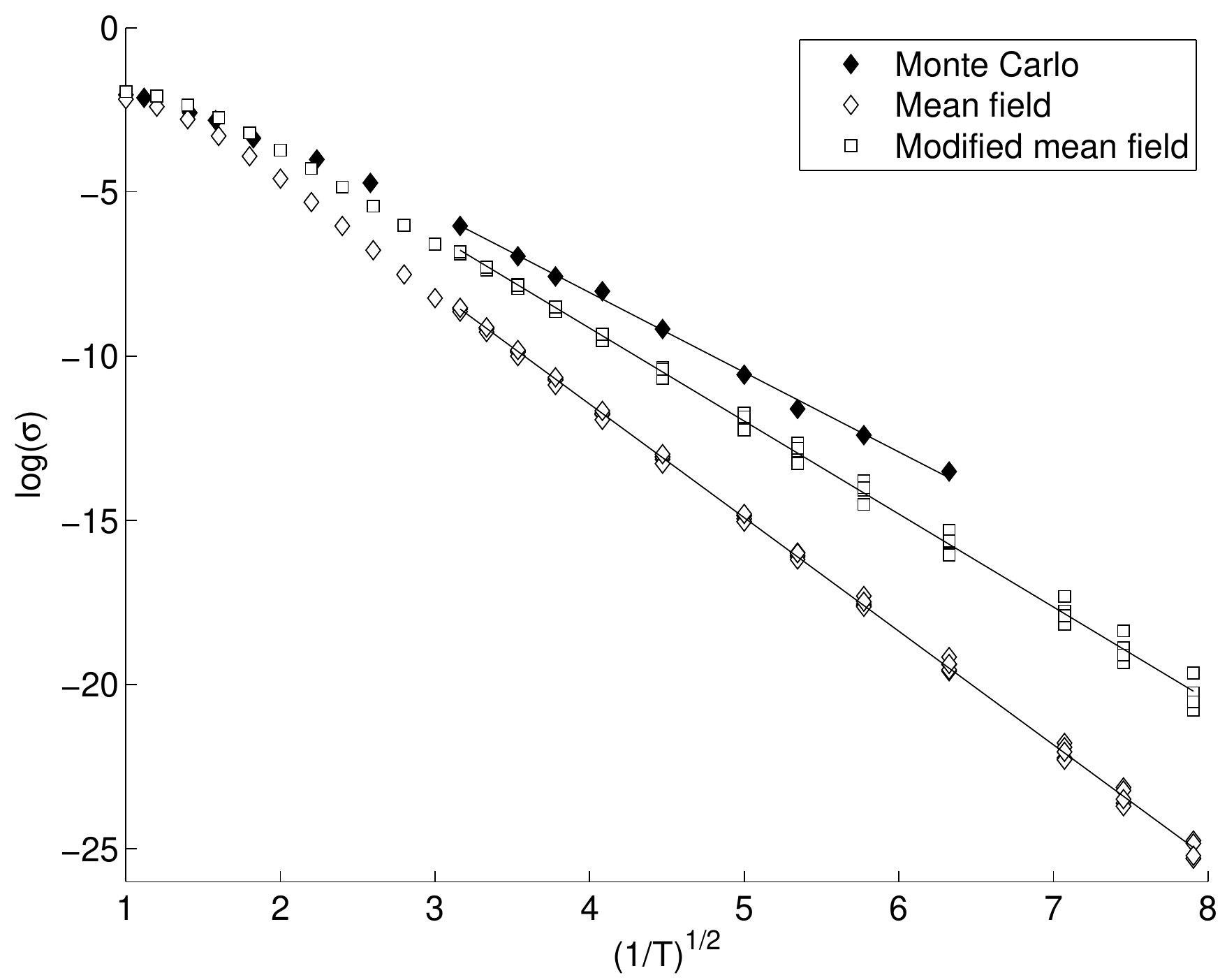}
\end{center}
\caption{\label{fig:cond} Logarithm of conductivity vs. $T^{-1/2}$  for Monte Carlo and mean field results, as specified in legend. Solid lines show linear fits at low temperatures, while markers are the results obtained from calculations/simulations.
  }
\end{figure}

The temperature dependence of the conductance is shown in
Fig.~\ref{fig:cond} for the three models. 
We see that all models follow closely the Efros-Shklovskii law
\eqref{eq:ES}, except at high temperatures where the conductance
becomes temperature independent. At high temperatures all three models
give close to the same conductance. At lower temperatures the mean
field approach underestimates the conductance. There are two reasons
for this. First, the mean field equations underestimates the number of
states close to the bottom of the Coulomb gap. This means that the
number of sites which are partially occupied, and active in transport
is underestimated. Second, the original mean field transition rates
are too small for an important class of transitions as discussed in
Sec. \ref{sec:rates}. The modified mean field rates corrects the
second problem, and we see that this brings the results into much
closer agreement with the Monte Carlo simulations. Assuming the
Efros-Shklovskii law to hold, we can extract $T_0$ from linear fits to
the data. We obtain  (in units of $e^2/\kappa a$)
$T_0= 5.9\pm0.4$ (Monte Carlo), $T_0=8.0\pm0.1$ (MLMF)
and $T_0=12.0\pm0.05$ (original LMF).   It turns out that
a self-consistent type of percolation approach based on the ESMF (see
Ref.~\onlinecite{levin1987} and references therein) produces good
values for $T_0$. In particular, for two-dimensional case $T_0 = 6.5
\pm 1$.\cite{Nguen84,Nguyen06} Note that the results for $T_0$ of the
MLMF and ESMF are significantly closer to those of the Monte Carlo
calculation than the results of the original LMF.   The
difference between the the MFLM and the ESMF is that the MFLM takes
into account (if only approximately, as discussed in
Sec. \ref{sec:equilibrium}) the smearing of the Coulomb gap at finite
temperature. This increases the density of states close to the Fermi
level, and we expect that it increases the conductance. This effect
should be more pronounced at higher temperatures, and therefore it is
natural that $T_0$ is larger when this effect is included, and this is
indeed what we see. Why the ESMF gives a value closer to the Monte
Carlo results is not clear, and it would be instructive to compare the
values of the conductance, not only $T_0$. In any case, the difference
is not very large, and we conclude that the smearing of the Coulomb
gap is not of great significance for DC conductance.

\section{Summary}\label{sec:discussion}

We have tested the recently proposed local mean field theory of the Coulomb
glass~\cite{amirMF,amirVRH} and compared it to Monte Carlo
simulations. We have also proposed a modified expression for the mean
field transition rates to take into account correlations at the most
basic level that a jump must take place from an empty to an occupied
site. 

We have found that the LMF equations underestimate the
number of sites in the Coulomb gap at finite temperatures. They will
also underestimate the number of sites with intermediate occupation
probability, resulting in occupation numbers that are close to either
0 (empty site) or 1 (filled site). 

The transition rates of the
original mean field equations are strongly underestimated for an
important class of transitions, where an electron jumps from below to
above the Fermi level, but the distance is short enough so that the
self interaction will compensate for the increase in SPE. 

The conductance follows closely the ES law in Monte Carlo
simulations and both the LMF calculations. However, the LMF gives values
of the current smaller than what is seen in the Monte Carlo
simulations. The difference is largest when using  the original LMF
expression for the rates. The MLMF rates come much closer to the
simulation results, indicating that a major part of the correlations
in the system can be understood in the simple pair approximation
used. More complicated correlations, involving three or more sites
certainly exist, but seem to be of less importance. The MLMF results are also close to what was found using the ESMF, indicating that the smearing of the Coulomb gap is not important in determining DC conductance.
This is not surprising since at very low temperatures, $T \ll T_0$, the typical energy band contributing to conductance, 
$\sim (TT_0)^{1/2}$, is much larger than temperature.  

This work is part of the master project of one of the authors (EB) and
more details can be found in his thesis.\cite{eivind}

\acknowledgements
We are grateful to B. I. Shklovskii and M. E. Raikh for critical remarks.


\begin{thebibliography}{35}%
\makeatletter
\providecommand \@ifxundefined [1]{%
 \@ifx{#1\undefined}
}%
\providecommand \@ifnum [1]{%
 \ifnum #1\expandafter \@firstoftwo
 \else \expandafter \@secondoftwo
 \fi
}%
\providecommand \@ifx [1]{%
 \ifx #1\expandafter \@firstoftwo
 \else \expandafter \@secondoftwo
 \fi
}%
\providecommand \natexlab [1]{#1}%
\providecommand \enquote  [1]{``#1''}%
\providecommand \bibnamefont  [1]{#1}%
\providecommand \bibfnamefont [1]{#1}%
\providecommand \citenamefont [1]{#1}%
\providecommand \href@noop [0]{\@secondoftwo}%
\providecommand \href [0]{\begingroup \@sanitize@url \@href}%
\providecommand \@href[1]{\@@startlink{#1}\@@href}%
\providecommand \@@href[1]{\endgroup#1\@@endlink}%
\providecommand \@sanitize@url [0]{\catcode `\\12\catcode `\$12\catcode
  `\&12\catcode `\#12\catcode `\^12\catcode `\_12\catcode `\%12\relax}%
\providecommand \@@startlink[1]{}%
\providecommand \@@endlink[0]{}%
\providecommand \url  [0]{\begingroup\@sanitize@url \@url }%
\providecommand \@url [1]{\endgroup\@href {#1}{\urlprefix }}%
\providecommand \urlprefix  [0]{URL }%
\providecommand \Eprint [0]{\href }%
\providecommand \doibase [0]{http://dx.doi.org/}%
\providecommand \selectlanguage [0]{\@gobble}%
\providecommand \bibinfo  [0]{\@secondoftwo}%
\providecommand \bibfield  [0]{\@secondoftwo}%
\providecommand \translation [1]{[#1]}%
\providecommand \BibitemOpen [0]{}%
\providecommand \bibitemStop [0]{}%
\providecommand \bibitemNoStop [0]{.\EOS\space}%
\providecommand \EOS [0]{\spacefactor3000\relax}%
\providecommand \BibitemShut  [1]{\csname bibitem#1\endcsname}%
\let\auto@bib@innerbib\@empty
\bibitem [{\citenamefont {Mott}(1968)}]{mott}%
  \BibitemOpen
  \bibfield  {author} {\bibinfo {author} {\bibfnamefont {N.}~\bibnamefont
  {Mott}},\ }\href@noop {} {\bibfield  {journal} {\bibinfo  {journal} {Journal
  of Non-Crystalline Solids}\ }\textbf {\bibinfo {volume} {1}},\ \bibinfo
  {pages} {1} (\bibinfo {year} {1968})}\BibitemShut {NoStop}%
\bibitem [{\citenamefont {Ambegaokar}\ \emph {et~al.}(1971)\citenamefont
  {Ambegaokar}, \citenamefont {Halperin},\ and\ \citenamefont
  {Langer}}]{ambegaokar}%
  \BibitemOpen
  \bibfield  {author} {\bibinfo {author} {\bibfnamefont {V.}~\bibnamefont
  {Ambegaokar}}, \bibinfo {author} {\bibfnamefont {B.~I.}\ \bibnamefont
  {Halperin}}, \ and\ \bibinfo {author} {\bibfnamefont {J.~S.}\ \bibnamefont
  {Langer}},\ }\href {\doibase 10.1103/PhysRevB.4.2612} {\bibfield  {journal}
  {\bibinfo  {journal} {Phys. Rev. B}\ }\textbf {\bibinfo {volume} {4}},\
  \bibinfo {pages} {2612} (\bibinfo {year} {1971})}\BibitemShut {NoStop}%
\bibitem [{\citenamefont {Shklovskii}\ and\ \citenamefont
  {Efros}(1971)}]{ES71}%
  \BibitemOpen
  \bibfield  {author} {\bibinfo {author} {\bibfnamefont {B.~I.}\ \bibnamefont
  {Shklovskii}}\ and\ \bibinfo {author} {\bibfnamefont {A.~L.}\ \bibnamefont
  {Efros}},\ }\href@noop {} {\bibfield  {journal} {\bibinfo  {journal} {Zh.
  Eksp. Teor. Fiz.}\ }\textbf {\bibinfo {volume} {60}},\ \bibinfo {pages} {867}
  (\bibinfo {year} {1971})},\ \bibinfo {note} {[Sov. Phys.-JETP \textbf{33},
  468 (1971)]}\BibitemShut {NoStop}%
\bibitem [{\citenamefont {Pollak}(1972)}]{Pollak72}%
  \BibitemOpen
  \bibfield  {author} {\bibinfo {author} {\bibfnamefont {M.}~\bibnamefont
  {Pollak}},\ }\href@noop {} {\bibfield  {journal} {\bibinfo  {journal} {J.
  Non-Crystal. Solids}\ }\textbf {\bibinfo {volume} {11}},\ \bibinfo {pages}
  {1} (\bibinfo {year} {1972})}\BibitemShut {NoStop}%
\bibitem [{\citenamefont {Shklovskii}(1972)}]{Shklovskii72}%
  \BibitemOpen
  \bibfield  {author} {\bibinfo {author} {\bibfnamefont {B.~I.}\ \bibnamefont
  {Shklovskii}},\ }\href@noop {} {\bibfield  {journal} {\bibinfo  {journal}
  {Zh. Eksp. Teor. Fiz.}\ }\textbf {\bibinfo {volume} {61}},\ \bibinfo {pages}
  {2033} (\bibinfo {year} {1972})},\ \bibinfo {note} {[Sov. Phys.-JETP
  \textbf{34} , 108 (1972)]}\BibitemShut {NoStop}%
\bibitem [{\citenamefont {Pollak}(1970)}]{Pollak70}%
  \BibitemOpen
  \bibfield  {author} {\bibinfo {author} {\bibfnamefont {M.}~\bibnamefont
  {Pollak}},\ }\href@noop {} {\bibfield  {journal} {\bibinfo  {journal}
  {Discuss. Faraday Soc.}\ }\textbf {\bibinfo {volume} {50}},\ \bibinfo {pages}
  {13} (\bibinfo {year} {1970})}\BibitemShut {NoStop}%
\bibitem [{\citenamefont {Pollak}(1971)}]{Pollak71}%
  \BibitemOpen
  \bibfield  {author} {\bibinfo {author} {\bibfnamefont {M.}~\bibnamefont
  {Pollak}},\ }\href@noop {} {\bibfield  {journal} {\bibinfo  {journal} {Proc.
  R. Soc. London, Ser. A}\ }\textbf {\bibinfo {volume} {325}},\ \bibinfo
  {pages} {383} (\bibinfo {year} {1971})}\BibitemShut {NoStop}%
\bibitem [{\citenamefont {Srinivasan}(1971)}]{Srinivasan}%
  \BibitemOpen
  \bibfield  {author} {\bibinfo {author} {\bibfnamefont {G.}~\bibnamefont
  {Srinivasan}},\ }\href {\doibase 10.1103/PhysRevB.4.2581} {\bibfield
  {journal} {\bibinfo  {journal} {Phys. Rev. B}\ }\textbf {\bibinfo {volume}
  {4}},\ \bibinfo {pages} {2581} (\bibinfo {year} {1971})}\BibitemShut
  {NoStop}%
\bibitem [{\citenamefont {Efros}\ and\ \citenamefont
  {Shklovskii}(1975)}]{ES75}%
  \BibitemOpen
  \bibfield  {author} {\bibinfo {author} {\bibfnamefont {A.~L.}\ \bibnamefont
  {Efros}}\ and\ \bibinfo {author} {\bibfnamefont {B.~I.}\ \bibnamefont
  {Shklovskii}},\ }\href@noop {} {\bibfield  {journal} {\bibinfo  {journal} {J.
  Phys. C}\ ,\ \bibinfo {pages} {L49}} (\bibinfo {year} {1975})}\BibitemShut
  {NoStop}%
\bibitem [{\citenamefont {Pikus}\ and\ \citenamefont
  {Efros}(1994)}]{PikusEfros}%
  \BibitemOpen
  \bibfield  {author} {\bibinfo {author} {\bibfnamefont {F.~G.}\ \bibnamefont
  {Pikus}}\ and\ \bibinfo {author} {\bibfnamefont {A.~L.}\ \bibnamefont
  {Efros}},\ }\href {\doibase 10.1103/PhysRevLett.73.3014} {\bibfield
  {journal} {\bibinfo  {journal} {Phys. Rev. Lett.}\ }\textbf {\bibinfo
  {volume} {73}},\ \bibinfo {pages} {3014} (\bibinfo {year}
  {1994})}\BibitemShut {NoStop}%
\bibitem [{\citenamefont {Shklovskii}\ and\ \citenamefont {Efros}(1984)}]{ES}%
  \BibitemOpen
  \bibfield  {author} {\bibinfo {author} {\bibfnamefont {B.~I.}\ \bibnamefont
  {Shklovskii}}\ and\ \bibinfo {author} {\bibfnamefont {A.~L.}\ \bibnamefont
  {Efros}},\ }\href@noop {} {\emph {\bibinfo {title} {Electronic properties of
  doped semiconductors}}}\ (\bibinfo  {publisher} {Springer, Berlin},\ \bibinfo
  {year} {1984})\BibitemShut {NoStop}%
\bibitem [{\citenamefont {Amir}\ \emph {et~al.}(2008)\citenamefont {Amir},
  \citenamefont {Oreg},\ and\ \citenamefont {Imry}}]{amirMF}%
  \BibitemOpen
  \bibfield  {author} {\bibinfo {author} {\bibfnamefont {A.}~\bibnamefont
  {Amir}}, \bibinfo {author} {\bibfnamefont {Y.}~\bibnamefont {Oreg}}, \ and\
  \bibinfo {author} {\bibfnamefont {Y.}~\bibnamefont {Imry}},\ }\href {\doibase
  10.1103/PhysRevB.77.165207} {\bibfield  {journal} {\bibinfo  {journal} {Phys.
  Rev. B}\ }\textbf {\bibinfo {volume} {77}},\ \bibinfo {pages} {165207}
  (\bibinfo {year} {2008})}\BibitemShut {NoStop}%
\bibitem [{\citenamefont {Amir}\ \emph {et~al.}(2011)\citenamefont {Amir},
  \citenamefont {Oreg},\ and\ \citenamefont {Imry}}]{AmirAnu}%
  \BibitemOpen
  \bibfield  {author} {\bibinfo {author} {\bibfnamefont {A.}~\bibnamefont
  {Amir}}, \bibinfo {author} {\bibfnamefont {Y.}~\bibnamefont {Oreg}}, \ and\
  \bibinfo {author} {\bibfnamefont {J.}~\bibnamefont {Imry}},\ }\href@noop {}
  {\bibfield  {journal} {\bibinfo  {journal} {Annu. Rev. Condens. Matter.
  Phys.}\ }\textbf {\bibinfo {volume} {2}},\ \bibinfo {pages} {235} (\bibinfo
  {year} {2011})}\BibitemShut {NoStop}%
\bibitem [{\citenamefont {Amir}\ \emph {et~al.}(2012)\citenamefont {Amir},
  \citenamefont {Oreg},\ and\ \citenamefont {Imry}}]{AmirPNAS}%
  \BibitemOpen
  \bibfield  {author} {\bibinfo {author} {\bibfnamefont {A.}~\bibnamefont
  {Amir}}, \bibinfo {author} {\bibfnamefont {Y.}~\bibnamefont {Oreg}}, \ and\
  \bibinfo {author} {\bibfnamefont {J.}~\bibnamefont {Imry}},\ }\href@noop {}
  {\bibfield  {journal} {\bibinfo  {journal} {PNAS}\ }\textbf {\bibinfo
  {volume} {109}},\ \bibinfo {pages} {1850} (\bibinfo {year}
  {2012})}\BibitemShut {NoStop}%
\bibitem [{\citenamefont {Amir}\ \emph {et~al.}(2009)\citenamefont {Amir},
  \citenamefont {Oreg},\ and\ \citenamefont {Imry}}]{amirVRH}%
  \BibitemOpen
  \bibfield  {author} {\bibinfo {author} {\bibfnamefont {A.}~\bibnamefont
  {Amir}}, \bibinfo {author} {\bibfnamefont {Y.}~\bibnamefont {Oreg}}, \ and\
  \bibinfo {author} {\bibfnamefont {Y.}~\bibnamefont {Imry}},\ }\href {\doibase
  10.1103/PhysRevB.80.245214} {\bibfield  {journal} {\bibinfo  {journal} {Phys.
  Rev. B}\ }\textbf {\bibinfo {volume} {80}},\ \bibinfo {pages} {245214}
  (\bibinfo {year} {2009})}\BibitemShut {NoStop}%
\bibitem [{\citenamefont {Levin}\ \emph
  {et~al.}(1982{\natexlab{a}})\citenamefont {Levin}, \citenamefont {Nguyen},\
  and\ \citenamefont {Shklovskii}}]{levin82}%
  \BibitemOpen
  \bibfield  {author} {\bibinfo {author} {\bibfnamefont {E.~I.}\ \bibnamefont
  {Levin}}, \bibinfo {author} {\bibfnamefont {V.~L.}\ \bibnamefont {Nguyen}}, \
  and\ \bibinfo {author} {\bibfnamefont {B.~I.}\ \bibnamefont {Shklovskii}},\
  }\href@noop {} {\bibfield  {journal} {\bibinfo  {journal} {Zh. Eksp. Theor.
  Fiz.}\ }\textbf {\bibinfo {volume} {82}},\ \bibinfo {pages} {1591} (\bibinfo
  {year} {1982}{\natexlab{a}})},\ \bibinfo {note} {[Sov. Phys. JETP
  \textbf{55}, 921 (1982)]}\BibitemShut {NoStop}%
\bibitem [{\citenamefont {Levin}\ \emph
  {et~al.}(1982{\natexlab{b}})\citenamefont {Levin}, \citenamefont {Nguyen},\
  and\ \citenamefont {Shklovskii}}]{levin82a}%
  \BibitemOpen
  \bibfield  {author} {\bibinfo {author} {\bibfnamefont {E.~I.}\ \bibnamefont
  {Levin}}, \bibinfo {author} {\bibfnamefont {V.~L.}\ \bibnamefont {Nguyen}}, \
  and\ \bibinfo {author} {\bibfnamefont {B.~I.}\ \bibnamefont {Shklovskii}},\
  }\href@noop {} {\bibfield  {journal} {\bibinfo  {journal} {Fiz. Tekh.
  Poluprov.}\ }\textbf {\bibinfo {volume} {16}},\ \bibinfo {pages} {815}
  (\bibinfo {year} {1982}{\natexlab{b}})},\ \bibinfo {note} {[Sov. Phys.
  Semicond. \textbf{16}, 523 (1982)}\BibitemShut {NoStop}%
\bibitem [{\citenamefont {Pastor}\ and\ \citenamefont
  {Dobrosavljevic}(1999)}]{pastor99}%
  \BibitemOpen
  \bibfield  {author} {\bibinfo {author} {\bibfnamefont {A.~A.}\ \bibnamefont
  {Pastor}}\ and\ \bibinfo {author} {\bibfnamefont {V.}~\bibnamefont
  {Dobrosavljevic}},\ }\href@noop {} {\bibfield  {journal} {\bibinfo  {journal}
  {Phys. Rev. Lett.}\ }\textbf {\bibinfo {volume} {83}},\ \bibinfo {pages}
  {4642} (\bibinfo {year} {1999})}\BibitemShut {NoStop}%
\bibitem [{\citenamefont {Vojta}(1993)}]{vojta93}%
  \BibitemOpen
  \bibfield  {author} {\bibinfo {author} {\bibfnamefont {T.}~\bibnamefont
  {Vojta}},\ }\href@noop {} {\bibfield  {journal} {\bibinfo  {journal} {Journ.
  Phys. A}\ }\textbf {\bibinfo {volume} {26}},\ \bibinfo {pages} {2883}
  (\bibinfo {year} {1993})}\BibitemShut {NoStop}%
\bibitem [{\citenamefont {M\"uller}\ and\ \citenamefont
  {Ioffe}(2004)}]{muller04}%
  \BibitemOpen
  \bibfield  {author} {\bibinfo {author} {\bibfnamefont {M.}~\bibnamefont
  {M\"uller}}\ and\ \bibinfo {author} {\bibfnamefont {L.~B.}\ \bibnamefont
  {Ioffe}},\ }\href@noop {} {\bibfield  {journal} {\bibinfo  {journal} {Phys.
  Rev. Lett.}\ }\textbf {\bibinfo {volume} {93}},\ \bibinfo {pages} {256403}
  (\bibinfo {year} {2004})}\BibitemShut {NoStop}%
\bibitem [{\citenamefont {M\"uller}\ and\ \citenamefont
  {Pankov}(2007)}]{muller07}%
  \BibitemOpen
  \bibfield  {author} {\bibinfo {author} {\bibfnamefont {M.}~\bibnamefont
  {M\"uller}}\ and\ \bibinfo {author} {\bibfnamefont {S.}~\bibnamefont
  {Pankov}},\ }\href@noop {} {\bibfield  {journal} {\bibinfo  {journal} {Phys.
  Rev. B}\ }\textbf {\bibinfo {volume} {75}},\ \bibinfo {pages} {144201}
  (\bibinfo {year} {2007})}\BibitemShut {NoStop}%
\bibitem [{\citenamefont {Pankov}\ and\ \citenamefont
  {Dobrosavljevi\'c}(2005)}]{pankov05}%
  \BibitemOpen
  \bibfield  {author} {\bibinfo {author} {\bibfnamefont {S.}~\bibnamefont
  {Pankov}}\ and\ \bibinfo {author} {\bibfnamefont {V.}~\bibnamefont
  {Dobrosavljevi\'c}},\ }\href@noop {} {\bibfield  {journal} {\bibinfo
  {journal} {Phys. Rev. Lett.}\ }\textbf {\bibinfo {volume} {94}},\ \bibinfo
  {pages} {046402} (\bibinfo {year} {2005})}\BibitemShut {NoStop}%
\bibitem [{\citenamefont {Surer}\ \emph {et~al.}(2009)\citenamefont {Surer},
  \citenamefont {Katzgraber}, \citenamefont {Zimanyi}, \citenamefont
  {Allgood},\ and\ \citenamefont {Blatter}}]{katzgraber09}%
  \BibitemOpen
  \bibfield  {author} {\bibinfo {author} {\bibfnamefont {B.}~\bibnamefont
  {Surer}}, \bibinfo {author} {\bibfnamefont {H.~G.}\ \bibnamefont
  {Katzgraber}}, \bibinfo {author} {\bibfnamefont {G.~T.}\ \bibnamefont
  {Zimanyi}}, \bibinfo {author} {\bibfnamefont {B.~A.}\ \bibnamefont
  {Allgood}}, \ and\ \bibinfo {author} {\bibfnamefont {G.}~\bibnamefont
  {Blatter}},\ }\href {\doibase 10.1103/PhysRevLett.102.067205} {\bibfield
  {journal} {\bibinfo  {journal} {Phys. Rev. Lett.}\ }\textbf {\bibinfo
  {volume} {102}},\ \bibinfo {pages} {067205} (\bibinfo {year}
  {2009})}\BibitemShut {NoStop}%
\bibitem [{\citenamefont {Goethe}\ and\ \citenamefont
  {Palassini}(2009)}]{goethe2009}%
  \BibitemOpen
  \bibfield  {author} {\bibinfo {author} {\bibfnamefont {M.}~\bibnamefont
  {Goethe}}\ and\ \bibinfo {author} {\bibfnamefont {M.}~\bibnamefont
  {Palassini}},\ }\href {\doibase 10.1103/PhysRevLett.103.045702} {\bibfield
  {journal} {\bibinfo  {journal} {Phys. Rev. Lett.}\ }\textbf {\bibinfo
  {volume} {103}},\ \bibinfo {pages} {045702} (\bibinfo {year}
  {2009})}\BibitemShut {NoStop}%
\bibitem [{\citenamefont {Tsigankov}\ and\ \citenamefont
  {Efros}(2002)}]{tsigankov}%
  \BibitemOpen
  \bibfield  {author} {\bibinfo {author} {\bibfnamefont {D.~N.}\ \bibnamefont
  {Tsigankov}}\ and\ \bibinfo {author} {\bibfnamefont {A.~L.}\ \bibnamefont
  {Efros}},\ }\href {\doibase 10.1103/PhysRevLett.88.176602} {\bibfield
  {journal} {\bibinfo  {journal} {Phys. Rev. Lett.}\ }\textbf {\bibinfo
  {volume} {88}},\ \bibinfo {pages} {176602} (\bibinfo {year}
  {2002})}\BibitemShut {NoStop}%
\bibitem [{\citenamefont {Tsigankov}\ \emph {et~al.}(2003)\citenamefont
  {Tsigankov}, \citenamefont {Pazy}, \citenamefont {Laikhtman},\ and\
  \citenamefont {Efros}}]{tsigankovPRB}%
  \BibitemOpen
  \bibfield  {author} {\bibinfo {author} {\bibfnamefont {D.~N.}\ \bibnamefont
  {Tsigankov}}, \bibinfo {author} {\bibfnamefont {E.}~\bibnamefont {Pazy}},
  \bibinfo {author} {\bibfnamefont {B.~D.}\ \bibnamefont {Laikhtman}}, \ and\
  \bibinfo {author} {\bibfnamefont {A.~L.}\ \bibnamefont {Efros}},\ }\href
  {\doibase 10.1103/PhysRevB.68.184205} {\bibfield  {journal} {\bibinfo
  {journal} {Phys. Rev. B}\ }\textbf {\bibinfo {volume} {68}},\ \bibinfo
  {pages} {184205} (\bibinfo {year} {2003})}\BibitemShut {NoStop}%
\bibitem [{\citenamefont {Voje}(2009)}]{aurora}%
  \BibitemOpen
  \bibfield  {author} {\bibinfo {author} {\bibfnamefont {A.}~\bibnamefont
  {Voje}},\ }\emph {\bibinfo {title} {Non-Ohmic Variable Range Hopping in
  Lightly Doped Semiconductors}},\ \href@noop {} {Master's thesis},\ \bibinfo
  {school} {UiO} (\bibinfo {year} {2009}),\ \bibinfo {note}
  {\url{http://urn.nb.no/URN:NBN:no-26080}}\BibitemShut {NoStop}%
\bibitem [{Note1()}]{Note1}%
  \BibitemOpen
  \bibinfo {note} {Note that MF calculations of DOS do not involve transition
  rates. Therefore, LMF and MLMF calculations of DOS are
  equivalent.}\BibitemShut {Stop}%
\bibitem [{\citenamefont {Glatz}\ \emph {et~al.}(2008)\citenamefont {Glatz},
  \citenamefont {Vinokur}, \citenamefont {Bergli}, \citenamefont {Kirkengen},\
  and\ \citenamefont {Galperin}}]{glatz2008}%
  \BibitemOpen
  \bibfield  {author} {\bibinfo {author} {\bibfnamefont {A.}~\bibnamefont
  {Glatz}}, \bibinfo {author} {\bibfnamefont {V.~M.}\ \bibnamefont {Vinokur}},
  \bibinfo {author} {\bibfnamefont {J.}~\bibnamefont {Bergli}}, \bibinfo
  {author} {\bibfnamefont {M.}~\bibnamefont {Kirkengen}}, \ and\ \bibinfo
  {author} {\bibfnamefont {Y.~M.}\ \bibnamefont {Galperin}},\ }\href
  {http://stacks.iop.org/1742-5468/2008/i=06/a=P06006} {\bibfield  {journal}
  {\bibinfo  {journal} {Journal of Statistical Mechanics: Theory and
  Experiment}\ }\textbf {\bibinfo {volume} {2008}},\ \bibinfo {pages} {P06006}
  (\bibinfo {year} {2008})}\BibitemShut {NoStop}%
\bibitem [{\citenamefont {Levin}\ \emph {et~al.}(1987)\citenamefont {Levin},
  \citenamefont {Nguen}, \citenamefont {Shklovskii},\ and\ \citenamefont
  {Efros}}]{levin1987}%
  \BibitemOpen
  \bibfield  {author} {\bibinfo {author} {\bibfnamefont {E.~I.}\ \bibnamefont
  {Levin}}, \bibinfo {author} {\bibfnamefont {V.~L.}\ \bibnamefont {Nguen}},
  \bibinfo {author} {\bibfnamefont {B.~I.}\ \bibnamefont {Shklovskii}}, \ and\
  \bibinfo {author} {\bibfnamefont {A.~L.}\ \bibnamefont {Efros}},\ }\href@noop
  {} {\bibfield  {journal} {\bibinfo  {journal} {Zh. Eksp. Teor. Fiz.}\
  }\textbf {\bibinfo {volume} {92}},\ \bibinfo {pages} {1499} (\bibinfo {year}
  {1987})},\ \bibinfo {note} {[Sov. Phys. JETP, \textbf{65}, 842
  (1987)]}\BibitemShut {NoStop}%
\bibitem [{\citenamefont {Mogilyanskii}\ and\ \citenamefont
  {Raikh}(1989)}]{raikh89}%
  \BibitemOpen
  \bibfield  {author} {\bibinfo {author} {\bibfnamefont {A.~A.}\ \bibnamefont
  {Mogilyanskii}}\ and\ \bibinfo {author} {\bibfnamefont {M.~{\'E}.}\
  \bibnamefont {Raikh}},\ }\href@noop {} {\bibfield  {journal} {\bibinfo
  {journal} {Zh. Eksp. Teor. Fiz.}\ }\textbf {\bibinfo {volume} {95}},\
  \bibinfo {pages} {1870} (\bibinfo {year} {1989})},\ \bibinfo {note} {[Sov.
  Phys. JETP, \textbf{68}, 1081 (1989)]}\BibitemShut {NoStop}%
\bibitem [{\citenamefont {Miller}\ and\ \citenamefont
  {Abrahams}(1960)}]{MillerAbrahams}%
  \BibitemOpen
  \bibfield  {author} {\bibinfo {author} {\bibfnamefont {A.}~\bibnamefont
  {Miller}}\ and\ \bibinfo {author} {\bibfnamefont {E.}~\bibnamefont
  {Abrahams}},\ }\href {\doibase 10.1103/PhysRev.120.745} {\bibfield  {journal}
  {\bibinfo  {journal} {Phys. Rev.}\ }\textbf {\bibinfo {volume} {120}},\
  \bibinfo {pages} {745} (\bibinfo {year} {1960})}\BibitemShut {NoStop}%
\bibitem [{\citenamefont {Nguen}(1984)}]{Nguen84}%
  \BibitemOpen
  \bibfield  {author} {\bibinfo {author} {\bibfnamefont {V.~L.}\ \bibnamefont
  {Nguen}},\ }\href@noop {} {\bibfield  {journal} {\bibinfo  {journal} {Fiz.
  Tekh. Poluprovodn.}\ }\textbf {\bibinfo {volume} {18}},\ \bibinfo {pages}
  {207} (\bibinfo {year} {1984})},\ \bibinfo {note} {[Sov. Phys. Semicond.
  \textbf{18}, 207 (1984)]}\BibitemShut {NoStop}%
\bibitem [{\citenamefont {Nguyen}\ \emph {et~al.}(2006)\citenamefont {Nguyen},
  \citenamefont {Nguyen},\ and\ \citenamefont {Dang}}]{Nguyen06}%
  \BibitemOpen
  \bibfield  {author} {\bibinfo {author} {\bibfnamefont {V.~D.}\ \bibnamefont
  {Nguyen}}, \bibinfo {author} {\bibfnamefont {V.~L.}\ \bibnamefont {Nguyen}},
  \ and\ \bibinfo {author} {\bibfnamefont {D.~T.}\ \bibnamefont {Dang}},\
  }\href@noop {} {\bibfield  {journal} {\bibinfo  {journal} {Phys. Lett. A}\
  }\textbf {\bibinfo {volume} {349}},\ \bibinfo {pages} {404} (\bibinfo {year}
  {2006})}\BibitemShut {NoStop}%
\bibitem [{\citenamefont {Bardalen}(2011)}]{eivind}%
  \BibitemOpen
  \bibfield  {author} {\bibinfo {author} {\bibfnamefont {E.}~\bibnamefont
  {Bardalen}},\ }\emph {\bibinfo {title} {Coulomb Glasses: A Comparison Between
  Mean Field and Monte Carlo Results}},\ \href@noop {} {Master's thesis},\
  \bibinfo  {school} {University of Oslo}, \bibinfo {address} {Norway}
  (\bibinfo {year} {2011}),\ \bibinfo {note}
  {\url{http://urn.nb.no/URN:NBN:no-28533}}\BibitemShut {NoStop}%
\end{thebibliography}

%

\end{document}